\documentclass[letter,12pt]{article}
\pdfoutput=1 % if your are submitting a pdflatex (i.e. if you have
\usepackage{jheppub} % for details on the use of the package, please
\usepackage[utf8]{inputenc}

\usepackage{mathtools}

\usepackage{epsfig}
\usepackage{amssymb}
\usepackage{amsfonts}
\usepackage{amsmath}
\usepackage{titlesec}
\usepackage{yfonts}

\usepackage{adjustbox}

\usepackage{arydshln}

\usepackage{ dsfont }

\usepackage[russian,english]{babel}
\usepackage[T1,T2A]{fontenc} % if needed

\usepackage[utf8]{inputenc}
\usepackage{amsmath}
\usepackage{calc}

\usepackage{accents}

\def\C{{\mathcal C}}
\def\c{{\rm c}}

\newcommand{\Z}{\ensuremath{\mathbb Z}}
\newcommand{\R}{\ensuremath{\mathbb R}}
\newcommand{\N}{\ensuremath{\mathbb N}}

\usepackage{amsmath}

\usepackage{ upgreek }
\usepackage{physics}

\def\bea{\begin{eqnarray}}
\def\eea{\end{eqnarray}}

\mathchardef\mhyphen="2D

\title{Comments on the holographic description of Narain theories}

\author[a,b]{Anatoly Dymarsky,} 
\author[a]{and Alfred Shapere} 
 
\affiliation[a]{Department of Physics and Astronomy, \\ University of Kentucky,\\ 505 Rose Street, Lexington, KY,  40506, USA\\}
\affiliation[b]{Skolkovo Institute of Science and Technology,\\Skolkovo Innovation Center, Moscow, Russia\\}
\emailAdd{a.dymarsky@uky.edu}
\emailAdd{shapere@g.uky.edu}

\abstract{
We discuss the holographic description  of Narain $U(1)^c\times U(1)^c$ conformal field theories, and their potential similarity to
conventional weakly coupled gravitational theories in the bulk, in the sense that the effective IR bulk description includes ``$U(1)$ gravity'' amended with additional  light degrees of freedom.  Starting from this picture, we formulate the hypothesis that in the large central charge limit the density of states of any Narain theory is bounded by below by the density of states of $U(1)$ gravity.  This immediately implies that the maximal value of the spectral gap for primary fields is $\Delta_1=c/(2\pi e)$. To test the self-consistency of this proposal, we study its implications using chiral lattice CFTs and CFTs based on quantum stabilizer codes. First we notice that the conjecture yields a new bound on quantum stabilizer codes, which is compatible with previously known bounds in the literature. We proceed to discuss the variance of the density of states, which for consistency must be vanishingly small in the large-$c$ limit. We consider ensembles of code and chiral theories and show that in both cases the density variance is exponentially small in the central charge.  
}

\begin{document} 
\maketitle
\flushbottom

\section{Introduction}
\label{sec:intro}

The conventional picture of the holographic correspondence equates quantum gravity in $AdS_3$ with a 2d  CFT at the boundary. In this correspondence, an individual conformal theory is dual to a particular theory of gravity, while many theories of gravity are expected to be similar in the IR. A particularly interesting problem would be to find the holographic dual to pure gravity in $AdS_3$, a hypothetical UV completion of the quasi-classical Einstein theory.  An attempt to evaluate the partition  function of this theory yields negative level degeneracies and a continuous spectrum \cite{Maloney:2007ud,witten2007three},  suggesting that it is not dual to any specific CFT. More recently these problems have been at least partially addressed by adding additional states to the bulk description \cite{Benjamin:2016aww,Alday:2019vdr,Benjamin:2020mfz,Alday:2020qkm}. This is consistent with the conventional expectation that certain large central charge $c\gg 1$ theories would  be described by a combination of quasi-classical gravity and additional matter fields.

Using modular invariance of the CFT partition function,  Hartman, Keller and  Stoica \cite{hartman2014universal} have shown that as long as these additional states contribute at subleading  order in $1/c$ to the entropy at large temperatures $\beta<2\pi$, the spectrum of CFT light states must be sparse 
\bea
\rho(\Delta) \leq e^{2\pi \Delta}, \qquad 0<\Delta \leq c/12. \label{sparseness}
\eea 
Conversely, if the sparseness condition \eqref{sparseness}  is satisfied, the density of states for $\Delta>c/6$ at leading order in $1/c$ is given by the Cardy formula 
\bea
\rho_{\rm Cardy}(\Delta)=e^{\pi \sqrt{4c/3 (\Delta-c/12)}},  \label{Cardyrho}
\eea
which matches the density of states of the BTZ black hole, suggesting that the gravity dual is a theory of quasiclassical gravity plus $o(c)$  additional matter fields. To reiterate, the conventional picture is that sparse 2d CFTs with large central charge are expected to be dual to quasiclassical gravity with some additional fields.

This simple picture immediately implies a bound on the maximal value of the spectral gap in large-$c$ CFTs, $\Delta_1\leq c/12$. Indeed, for $\Delta_1 \geq c/12$, the theory is automatically sparse. Once we assume it is dual to quasiclassical gravity, no additional matter can prevent black hole formation at  energies exceeding the BTZ threshold $E=\Delta-c/12>0$. Thus for $\Delta>c/12$ there must be CFT states dual to  black hole microstates and one obtains $\rho(\Delta)\geq \rho_{\rm Cardy}>0$ for $\Delta>c/12$ in sparse large-$c$ theories.  We emphasize that this picture is an unproven  but widely expected hypothesis.   

An indirect validation of this conventional picture could come from an independent proof that $\Delta_1\leq c/12$ in large $c$ theories. This problem has been attacked using the modular conformal bootstrap approach, starting with the bound set by Hellerman $\Delta_1\leq c/6 + O(1)$ \cite{Hellerman,qualls2014bounds,collier2018modular,hartman2019sphere}. So far the best asymptotic bound $\Delta_1\lesssim c/9.1$ \cite{afkhami2019fast} falls short of the desired value, which could indicate either a flaw in the general picture or the  relevance of other CFT consistency conditions beyond modular invariance to ensure a tight bound on the spectral gap. 

Recently a few novel examples of the holographic correspondence in the context of JT gravity \cite{Saad:2019lba} and Narain CFTs \cite{afkhami2020free,maloney2020averaging}, suggest that gravity in the bulk is dual not to a particular boundary theory, but to an average over an ensemble of such theories, 
{see \cite{belin2020random,cotler2020ads,maxfield2020path,cotler2020ads1} for  related developments in 3D gravity}. In retrospect this may explain the continuous spectrum of pure gravity in  $AdS_3$, which could arise from such an average.  At the same time there are other examples of holographic correspondences, starting from the original ${\mathcal N=4}$ SYM in the 4d dual to IIB SUGRA on $AdS_5 \times S_5$, which seem to leave no room for an ensemble interpretation. 

In our opinion these two scenarios for holographic correspondences are not inconsistent. Rather we will advocate a scenario in which each individual CFT is dual to some theory of quantum gravity in the bulk, and moreover, averaging over particular ensembles of boundary theories may also have a local description in the bulk.   The goal of this paper is to discuss the consistency of this picture for Narain theories --- namely, that individual Narian CFTs can be described as ``pure'' $U(1)$ gravity plus additional matter fields, where by pure $U(1)$ gravity  (in the notation of \cite{afkhami2020free}) we mean the  perturbative sector of $U(1)^c\times U(1)^c$ Chern-Simons theory in the bulk, which is dual to the average over all Narain theories. We repeat the analysis of Hartman, Keller and  Stoica for Narain theories and find the sparseness condition to be 
\bea
\varrho(\Delta)\leq e^{2\pi \Delta},\qquad 0<\Delta \leq c/(4\pi),
\eea
where by $\varrho(\Delta)$ we understand the density of $U(1)^c \times U(1)^c$ primary states. Provided this condition is satisfied, at leading  order in $1/c$ the density of states is then given by the analog of the Cardy formula \cite{afkhami2020high}
\bea
N(\Delta) \approx {(2\pi\, \Delta)^c \over \Gamma(c+1)},\qquad N(\Delta):=\int_0^\Delta d\Delta\, \varrho(\Delta), \label{Cardy}
\eea
for all $\Delta\geq c/(2\pi)$. As in  the case of  conventional gravity in $AdS_3$, in the case of $U(1)$ gravity the Cardy formula \eqref{Cardy} is valid 
for  $\Delta\gg 1$. For consistency of simplest picture, where in the large-$c$ limit any individual Narain CFT is described in the IR by $U(1)$ gravity plus additional matter, we require that the density of states of the Narain CFT should satisfy 
\bea
\label{conjecture}
{1\over c}\ln N(\Delta) \geq \ln (2\pi e\, \alpha),\qquad \alpha=\Delta/c>0, %
 \qquad c\rightarrow \infty. 
\eea
(This inequality should be understood strictly in the $c\rightarrow \infty$ limit.) 
In particular, from this condition follows a bound on the spectral gap for Narain theories, $\Delta_1\leq c/(2\pi e)$. To probe the consistency of this scenario and the hypothesis \eqref{conjecture}, we consider the code CFTs of \cite{Dymarsky:2020bps,Dymarsky:2020qom}  associated with quantum stabilizer codes as well as chiral theories associated with even self-dual lattices.   We show that \eqref{conjecture} leads to a novel bound on quantum codes, which is consistent with previously known results. Further, consistency of \eqref{conjecture} requires the mean value to be greater than or equal to all values in the ensemble, which is only possible  if the variance (and higher moments) vanish in the large-$c$ limit. We calculate the variance of $\ln N$ for the ensembles of code and chiral  theories and find it to be exponentially small $e^{-O(c)}$.

This paper is organized as follows. In the following section we remind the reader about the basics of the Narain theories and the duality between the average over Narain theories and $U(1)$ gravity in the bulk. We also introduce code CFTs and calculate their density of states in terms of the underlying quantum code. Section \ref{sec:HKS} repeats the analysis of 
Hartman, Keller and  Stoica for the Narain theories. We also formulate conjecture \eqref{conjecture} there. We use code and chiral theories to probe the consistency of the conjecture in section \ref{sec:analysis}. Section   \ref{sec:discussion}  concludes the paper with a discussion of implications for quantum gravity and the sphere packing problem. 

\section{Preliminaries}
\subsection{Averaged Narain theories and $U(1)$ gravity}
A Narain CFT describes $c \in \N$ free scalar fields compactified on a $c$-dimensional torus. 
Mathematically such  a theory  is parametrized by an even self-dual lattice $\Lambda$  in $\R^{c,c}$. 
The CFT torus partition function is given by 
\bea
Z_\Lambda(\tau,\bar \tau)={\Theta_{\Lambda}(\tau,\bar \tau)\over |\eta(\tau)|^{2c}},
\eea
where the Siegel-Narain theta function is 
\bea
\Theta_{\Lambda}=\sum_{(p_L,p_R)\in \Lambda} q^{p_L^2/2}\, {\bar q}^{p_R^2/2},\qquad q=e^{2\pi i\tau},\qquad {\bar q}=e^{-2\pi i \bar \tau}.
\eea

Narain theories exhibit a $U(1)^c \times U(1)^c$ symmetry. By density of states $\varrho(\Delta)$ we will understand density of $U(1)^c \times U(1)^c$ primaries, i.e.~points of the Narain lattice. 
Instead of the density $\varrho(\Delta)$ it is often convenient to study the cumulative number of states with dimensions bounded by some $\Delta$, which is the same as the number of points of $\Lambda$ inside a sphere of radius $R$, where $R^2/2=\Delta$: 
\bea
N(\Delta)=\int_0^\Delta \varrho(\Delta) d\Delta =
  \sum_{\substack{(p_L,p_R)\in \Lambda, \\[2pt] p_L^2+p_R^2\leq 2 \Delta}} 1.
\eea
Since $\Lambda$ is unimodular, each lattice cell has unit volume and therefore for sufficiently large $\Delta$ the cumulative number of states is equal to the volume of a $2c$-dimensional sphere of radius $R$,
\bea
\label{volume}
N_{\rm C}(\Delta) \approx {(2\pi\, \Delta)^c \over \Gamma(c+1)},\qquad \ln N_C(\Delta) \approx {c}\ln (2\pi e \Delta/c).
\eea
This is the analog of the Cardy formula for Narain theories, as can be deduced directly from modular invariance \cite{afkhami2020high}. It applies universally  for $\Delta \gg c$ but as we will see below, in certain cases its validity extends to much smaller values of $\Delta$. 
In what follows we will work in  large $c$ limit and introduce 
\bea
\lambda(\alpha):=\lim_{c\rightarrow \infty} {\ln N(\alpha c)\over c}=\lim_{c\rightarrow \infty} {\ln \varrho(\alpha c)\over c},
\eea
where the limit is taken with respect to a family of theories defined for arbitrarily large $c$. In this notation the Cardy formula is simply 
\bea
\lambda_{C}(\alpha)=\ln(2\pi\, e\, \alpha). \label{Cl}
\eea

We advocate the point of view that each Narain theory, characterized by a lattice $\Lambda$, is dual to some gauge theory in the bulk. We leave the details of the bulk microscopic description to future work, and here only discuss provisional properties of such theories in the IR, namely the possible behavior of $\lambda(\alpha)$. 

It has been argued recently that averaging over the ensemble of all Narain theories is dual to ``$U(1)$ gravity'' -- the pertubative sector of $U(1)^c\times U(1)^c$  Chern-Simons theory in the bulk. Because of averaging,  density of states is a continuous function \cite{afkhami2020free}
\bea
\varrho_{U(1)\, \it grav}(\Delta)=\delta(\Delta)+\sum_{|\ell|\leq \Delta} {2\pi^c \sigma_{1-c}(\ell)\over \Gamma(c/2)^2\zeta(c)}(\Delta-\ell^2)^{c/2-1}.
\eea
 For $\Delta\gg 1$ the summation over $\ell$ can be replaced by an integration, yielding  \eqref{volume} in the $c\gg 1$ limit. Hence, for this $U(1)$ gravity theory, the Cardy formula \eqref{Cl} is valid  for all values of $\Delta >c/(2\pi e)$, i.e.~when the number of states is exponential.  This mirrors the behavior of sparse holographic theories with Virasoro symmetry \cite{hartman2014universal}, see below.  

To conclude this section we  introduce chiral theories associated with even self-dual lattices in $\R^c$. In this case $c$ is divisible by $8$ and it is convenient to introduce $k=c/2$. For a given lattice $\Uplambda\in \R^c$ the partition function is 
\bea
Z_\Uplambda(\tau)={\Theta_\Uplambda(\tau)\over \eta(\tau)^c},
\eea
where lattice theta-series is
\bea
\Theta_\Uplambda(\tau)=\sum_{v\in \Uplambda} q^{v^2/2},\qquad q=e^{2\pi i \tau}.
\eea
As in the Narain case the Cardy formula is given by the volume of a $c$-dimensional sphere 
\bea
N(\Delta)={(2\pi\Delta)^k\over \Gamma(k+1)}, \label{Cardychiral}
\eea
which applies to any theory for $\Delta\gg k$. Averaging over the ensemble of all even self-dual lattices (with the weight specified by the size of the lattice automorphism group) yields \cite{nebe2006self}
\bea
\overline \Theta=E_k(\tau),
\eea
where $E_k$ is the Eisenstein series. By definition it has a representation as a sum over $\Gamma_\infty \backslash {\rm SL}(2,\Z)$, 
\bea
\overline{Z}(\tau)=\sum_{\gamma\in \Gamma_{\infty}\backslash {\rm SL}(2,\Z)} {1\over \eta^n(\gamma \tau)},  \label{holographychiral}
\eea
which can be interpreted as a sum over different handlebodies on the gravity side. This led to a suggestion in \cite{Dymarsky:2020bps} that the ensemble of chiral theories also has a holographic description, similar to the one of the Narain case. The Fourier expansion of the Eisenstein series is 
\bea
E_k=1+{2\over \zeta(1-k)}\sum_m \sigma_{k-1}(m) q^m, \label{E}
\eea
where the divisor function $\sigma_{k-1}(m)=\sum_{d|m} d^{k-1}$.
For large $k\gg 1$, $\sigma_{k-1}(m)=m^{k-1}\left(1+O(e^{-O(k)})\right)$ and 
\bea
{2\over \zeta(1-k)}={(2\pi)^k \over \Gamma(k)\zeta(k)}= {(2\pi)^k \over \Gamma(k)}\left(1+O(e^{-O(k)})\right)
\eea
leading to the density of states (dropping exponentially small corrections)
\bea
\overline{\varrho}(\Delta)=\delta(\Delta)+ {(2\pi )^k  \Delta^{k-1} \over \Gamma(k)}. \label{densitychiral}
\eea
 In the chiral case $\Delta$ assumes only integer values, i.e.~the density of states is a sum of delta-functions. When $\Delta\gg 1$ we can approximate it by a continuous distribution by treating  $\Delta$ in \eqref{densitychiral} as a continuous variable. 

Clearly \eqref{densitychiral} agrees with \eqref{Cardychiral}, which means that the Cardy formula applies for $\Delta \gg 1$. 
The number of states grows exponentially for $\Delta>c/(4\pi e)$. By extending the conjecture  \eqref{conjecture} to chiral theories, we can predicts the spectral gap in large $c$ limit to be bounded by $\Delta_1\leq c/(4\pi e)$.

\subsection{Density of states of code CFTs}
\label{sec:codeCFTs}
Code CFTs comprise a family of Narain CFTs associated with quantum stabilizer codes, which was  introduced in \cite{Dymarsky:2020bps,Dymarsky:2020qom}. Quantum codes are $c$-dimensional linear subspaces of $\Z_2^{2c}$ over $\Z_2$, self-orthogonal with respect to the metric
\bea
\label{g}
g=\left(
\begin{array}{cc}
 0 & I \\
 I & 0 \\
\end{array}
\right).
\eea
 A code $\C$ is non-uniquely characterized by its refined enumerator 
\bea
W_{\C}(x,y,z)=\sum_{\c\in \C} x^{n-w(\c)} y^{w_y{\c}} z^{w(\c)-w_y(\c)},\\
w(\c)=1^T\cdot c,\qquad w_y(\c)=(\c^T g\, \c)/2, \label{REP}
\eea
where $\vec{\, 1}$ is a vector of ones of length $2c$. The sum in \eqref{REP} goes over all $2^c$ codewords of the code $\C$.
Torus partition function of a code CFT is given by 
\bea
Z_\C(\tau,\bar \tau)={W_\C\left({b\,{\bar b}+c\,{\bar c}},{b\,{\bar b}-c\, {\bar c}},{a\, {\bar a}}\right)\over 2^n |\eta(\tau)|^{2n}}, \label{Z}
\eea
where 
\bea
a=\theta_2(\tau),\quad b=\theta_3(\tau),\quad c=\theta_4(\tau).
\eea
For pure imaginary $\tau=-\bar \tau=i\beta/2\pi$ corresponding Narain lattice theta function reduces to 
\bea
\Theta_\Lambda(\beta)=W_\C\left(\theta_3(q^2)^2,\theta_2(q^2)^2,\theta_3(q^2)\theta_2(q^2)\right).
\eea

\begin{figure}[t]
\begin{center}
\includegraphics[width=0.8\textwidth]{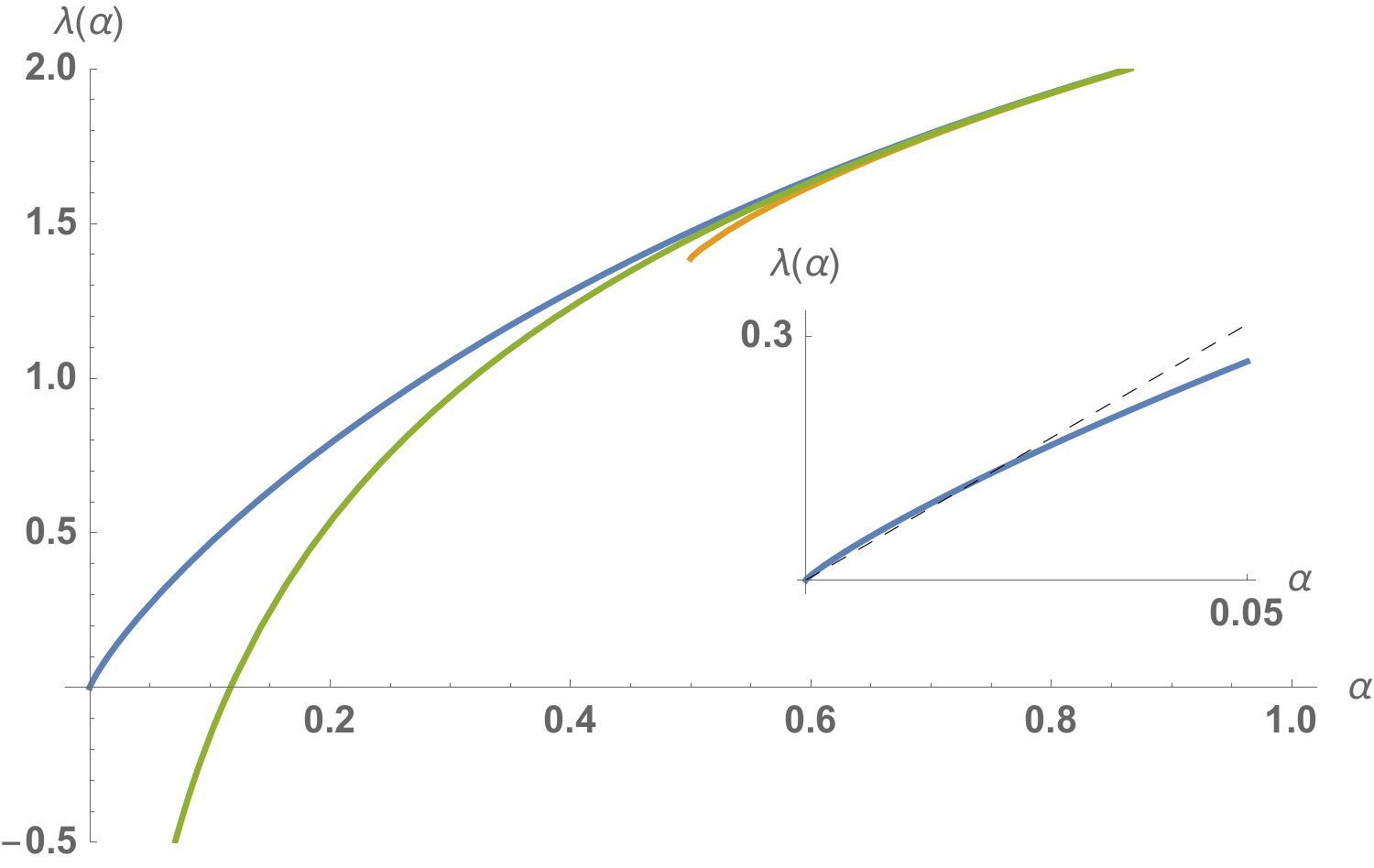}
\end{center}
\caption{Number of points of a cubic ``lattice'' of size $\sqrt{2}$ within a ball of radius $2\alpha c$. Blue line: 
$\lambda(\alpha,0)$. Orange line: $\lambda(\alpha,2)$. Green line:  $\ln(\pi e\alpha)$. For $\alpha<p/4$, formally $\lambda(\alpha,p)=-\infty$, which means no lattice vectors lie within the sphere of radius $2\alpha c < p c/2$. Inset: $\lambda(\alpha,0)$, blue line, vs sparseness condition $2\pi \alpha$, black dashed line, see section \ref{sec:HKS}.}
\label{fig:MO}
\end{figure}

The Narain lattice of a code CFT is a union of $2^c$ cubic ``lattices'' associated with the codewords, each being a conventional cubic lattice of the size $\sqrt{2}$ with the origin shifted to be at $\c/\sqrt{2}$ for some $\c\in \C$. For each such cubic ``lattice'' the number of points inside a sphere of radius $R^2=2\alpha c$ in the $c\rightarrow \infty$ limit was calculated by Mazo and Odlyzko \cite{mazo1990lattice} (see Appendix \ref{app:A} for a simple derivation),
\bea
N(\Delta,\c) &\approx& e^{c\, \lambda +O(c^{1/2})},\qquad \Delta=\alpha\, c,\qquad p=w(\c)/c, \\
\lambda(\alpha,p)&=&\lambda(\alpha,p,q^*), \label{MO1}
\eea
where $q^*$ is chosen to minimize the value of 
\bea
\label{MO2}
\lambda(\alpha,p,q)=-\alpha \ln(q)+(2-p)\ln \theta_3 (q^2)+ p \ln \theta_2(q^2),\quad \left.{\partial \lambda\over \partial q}\right|_{q^*}=0.
\eea
Depending on the value of $0\leq p\leq 2$, $\lambda(\alpha,p)$ approaches $\ln(\pi e \alpha)$ from above or below for large $\alpha$. Strictly speaking $\lambda(\alpha,p)$ always differs from $\ln(\pi e  \alpha)$, but convergence is very fast as can be see in Fig.~\ref{fig:MO}.
Summing over individual codewords and leaving leading exponent would give $\lambda$ for a code CFT. 

\begin{figure}[t]
\begin{center}
\includegraphics[width=0.8\textwidth]{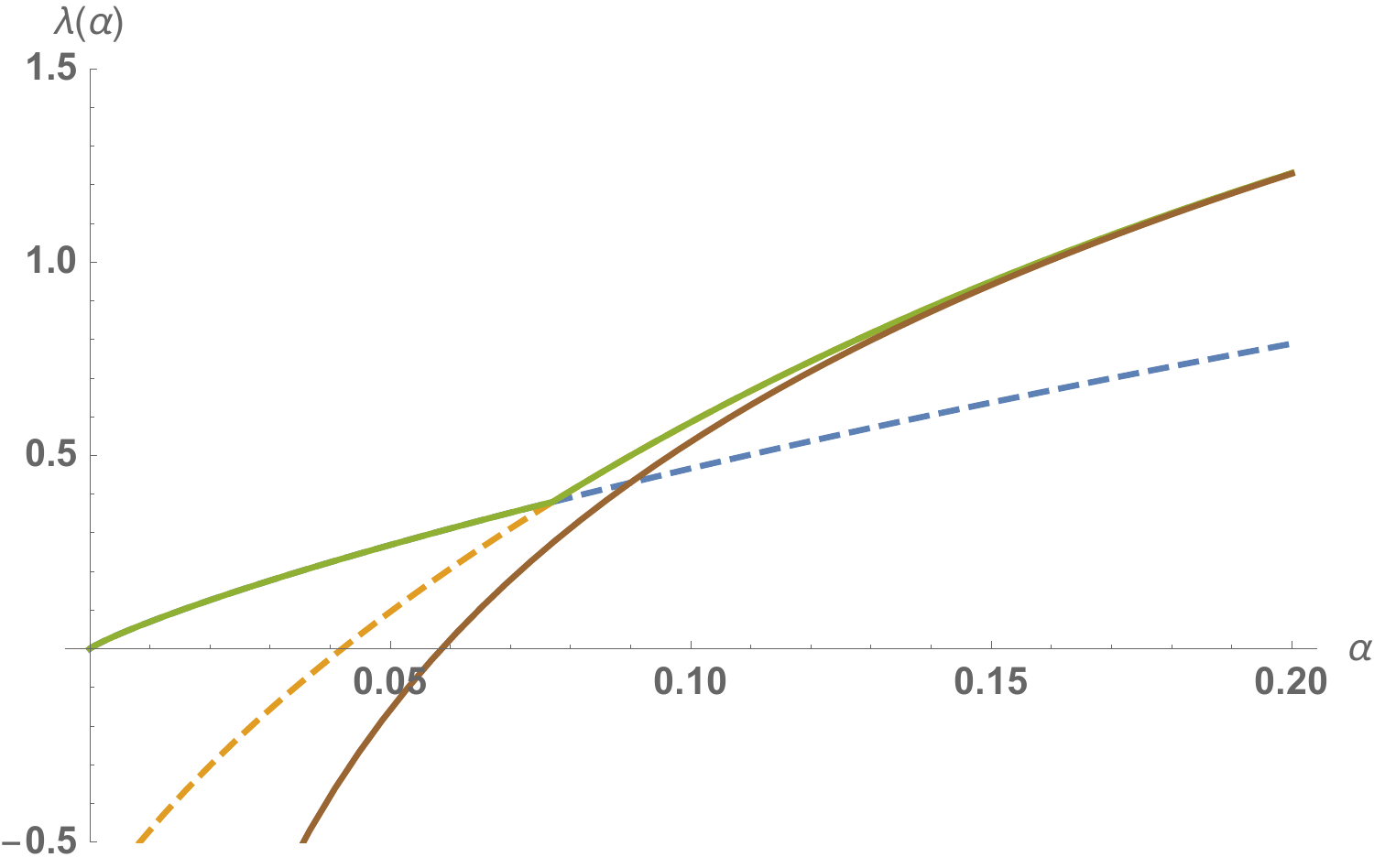}
\end{center}
\caption{Density of primary states of averaged code theory. Dashed blue and orange lines: ${\overline \lambda}_1$ and ${\overline \lambda}_2$ correspondingly. Green line: ${\overline \lambda}=\max({\overline \lambda}_1,{\overline \lambda}_2)$. Brown line: Cardy formula $\lambda_C=\ln(2\pi\, e\, \alpha)$.}
\label{fig:GV}
\end{figure}

As an example we consider averaging over the ensemble of all code CFTs, characterized by the averaged enumerator polynomial (see Appendix \ref{app:B})
\bea
\label{GV}
\overline{W}(1,t^2,t)=
1+{(1+t)^{2c}\over 2^c}+{(1+2t-t^2)^{c}\over 2^c}-2{(1+t)^c\over 2^c}.
\eea
For small $\alpha$ the first term (zero codeword) dominates, 
${\overline \lambda}_1=\lambda(\alpha,0)$.
For large $\alpha$ the second term is dominant, yielding 
for the theta-function
\bea
\Theta\approx {(\theta_3(q^2)+\theta_2(q^2))^{2c}\over 2^c}.
\eea
From here we find (see Appendix~\ref{app:A})
\bea
\label{lambda2}
{\overline \lambda}_2=-\alpha \ln q+2\ln(\theta_3(q^2)+\theta_2(q^2))-\ln(2),
\eea
where $q$ is chosen to minimize ${\overline \lambda}_2$. 
The third and fourth terms are not dominant for any value of $\alpha$.  The resulting density of states ${\overline \lambda}=\max ({\overline \lambda}_1,{\overline \lambda}_2)$
is shown in Fig.~\ref{fig:GV}.  
We expect the ensemble of code theories to exhibit the same important features as the full Narain ensemble, in particular self-averaging. Furthermore this ensemble may have a holographic interpretation as we argued in \cite{Dymarsky:2020qom}.

\section{HKS analysis for Narain  theories}
\label{sec:HKS}
Using modular invariance, Hartman, Keller, and Stoica have shown that as long as the full density of states of a 2d CFT is sparse  \eqref{sparseness}, the Cardy formula \eqref{Cardyrho} applies for any $\Delta> c/6$ in the large-$c$ limit. Equivalently, the free energy of a sparse large-$c$ theory for $\beta<2\pi$  is given at leading in $1/c$ order by 
\bea
\label{HKSF}
\ln Z=\left\{ \begin{array}{cc}
{c\beta \over 12}, &\quad  \beta>2\pi\\
{(2\pi)^2\over 12 \beta}, &\quad  \beta<2\pi
\end{array}
\right.,\qquad \tau=i \beta/2\pi.
\eea
Before proceeding to establish an analogous result for Narain theories, we note that no Narain theory is sparse in the sense of \eqref{sparseness}. Indeed, for large $\beta\gg 2\pi$ the theta function can be approximated as $\Theta_\Lambda=1+O(e^{-\Delta_1 \beta})$, where $\Delta_1$ is the dimension of the lightest non-trivial $U(1)^c \times U(1)^c$ primary.  Therefore in this limit 
\bea
\label{NZ}
\ln Z=-2c\ln(\eta(i\beta/(2\pi)))+O(e^{-\Delta_1 \beta})={c\,\beta\over 12}+2c\, e^{-\beta}+\dots 
\eea
This differs from \eqref{HKSF} by an amount that is exponentially small in $\beta$, but not by a $1/c$-suppressed factor.\footnote{The factor $O(e^{-\Delta_1 \beta})$ coming from the theta function is manifestly non-negative and therefore cannot cancel $2c\, e^{-\beta}$.} The exponentially small discrepancy between  \eqref{HKSF} and \eqref{NZ} at large $\beta$ implies that for $\Delta\gtrsim c$ the density of states would be described by the Cardy formula  \eqref{Cardyrho} to exponential precision.  For concreteness we focus on the averaged Narain theory ($U(1)$ gravity) for which, at leading order, 
\bea
\label{thetagrav}
\Theta_{U(1)\, \it grav}=
1+\left({2\pi \over \beta}\right)^c.
\eea
This expression is manifestly covariant under $\beta \rightarrow (2\pi)^2/\beta$. For $\beta>2\pi$ the free energy of $U(1)$ gravity is dominated by (descendants of) the vacuum state. This is the contribution of thermal $AdS_3$ in the bulk. For $\beta<2\pi$ second term dominates -- this is the contribution of  BTZ black hole geometry in the $U(1)$ gravity theory. At $\beta=2\pi$ both contributions are equal, marking the Hawking-Page transition.
 
\begin{figure}[t]
\begin{center}
\includegraphics[width=0.8\textwidth]{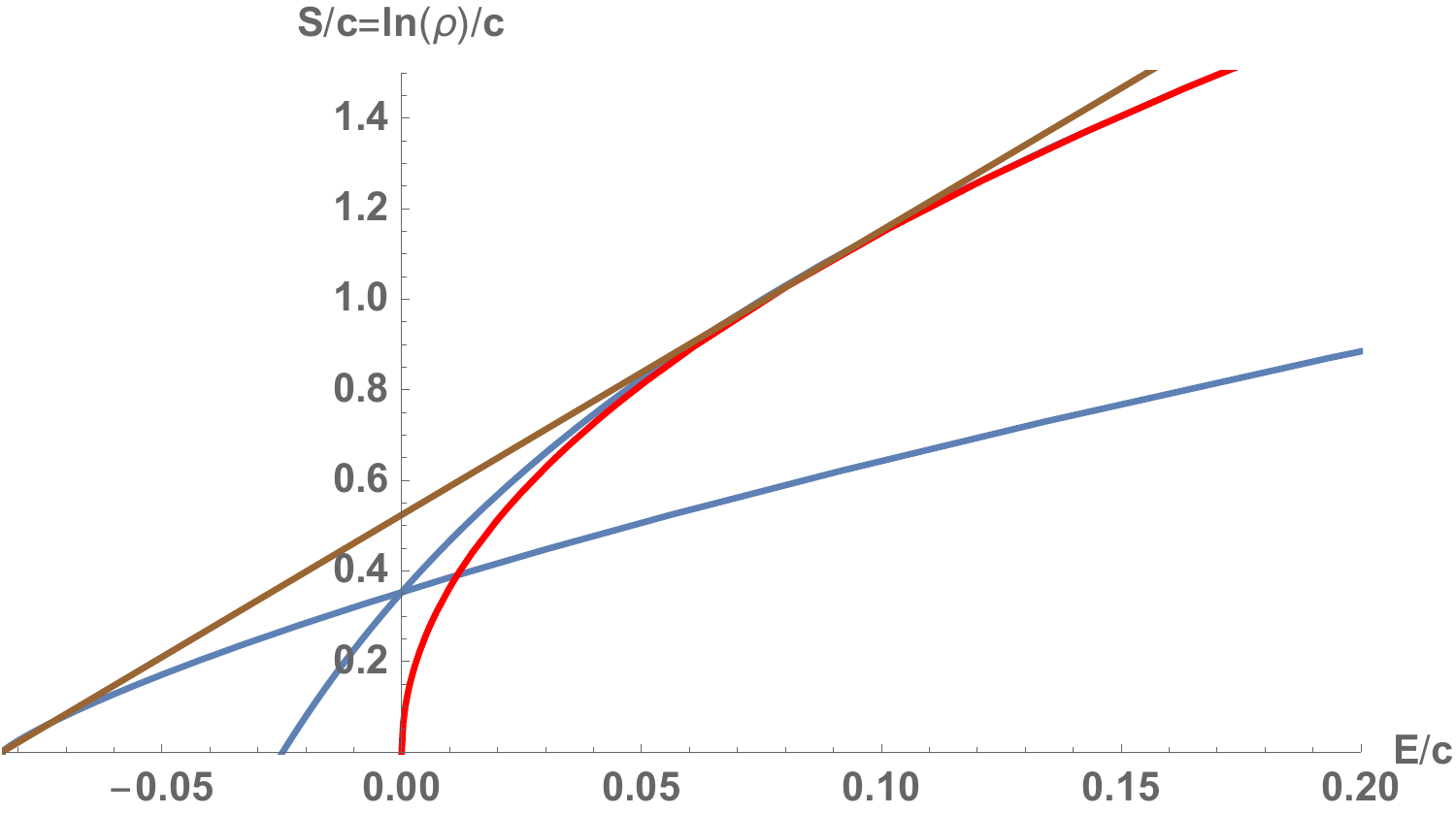}
\end{center}
\caption{Full density of states of the averaged Narain theory -- $U(1)$ gravity. 
Blue lines:  $S_1/c$ and $S_2/c$ as functions of $e=E/c$, where the full density of states (entropy) of $U(1)$ gravity is $S=\max(S_1,S_2)$.
$S_1$ dominates for $E<0$ and $S_2$ for $E>0$. 
Brown line: the sparseness condition $2\pi(e+1/12)$. $S_1(e)/c$ violates sparseness condition at small $e+1/12$. Red line: Cardy formula $\ln(\varrho)/c=\pi\sqrt{4e/3}$ \eqref{Cardyrho}.}
\label{fig:gravdensity}
\end{figure}

 We plot the entropy 
\bea
S=\ln Z+\beta E
\eea
as a function of $E=-\partial \ln Z/\partial \beta$ for
\bea
\ln Z_1/c=
-2\ln \eta(i\beta/2\pi),\qquad \ln Z_2/c=
-\ln(\beta/2\pi)-2 \ln \eta(i\beta/2\pi)
\eea
in Fig.~\ref{fig:gravdensity}, versus the Cardy formula \eqref{Cardyrho} and the HKS sparseness condition \eqref{sparseness}. For given $E$ the largest of $S_1,S_2$ is the leading contribution to the entropy. They have the same value for $E=0$. It is clear that the full density of states $S=S_2(E)$ is numerically very close to the Cardy formula for $\Delta\gtrsim c$, but in fact never actually matches it. This is consistent with  $S_1\approx \Delta \ln(\Delta/2c)$  for small $\Delta=(E+c/12)>0$, violating the sparseness condition $S\leq 2\pi \Delta$ \eqref{sparseness}.
As a final comment, we note that $S_2$ becomes non-negative at $\Delta=c/(2\pi e)$, which is exactly the threshold for new primary states, c.f.~\eqref{Cl}.

From now on we will focus on the density of primary states for a given Narain theory. Following HKS we split the theta series into two terms 
\bea
\label{modularc}
\Theta_\Lambda(\beta)&=&\Theta_L+\Theta_H,\, \, \, \, \quad \Theta_\Lambda(\beta')=\Theta_\Lambda(\beta)\left({ \beta\over 2\pi}\right)^c,\quad \beta'={(2\pi)^2\over \beta},\\
\Theta_L&=&\sum_{\Delta < \varepsilon} e^{-\beta \Delta},\qquad \Theta_H=\sum_{\Delta \geq  \varepsilon} e^{-\beta \Delta},
\eea
where sum is over primaries. 
Then for $\beta>2\pi$
\bea
\label{main}
\Theta_H(\beta)=
\sum_{\Delta \geq \varepsilon} e^{(\beta'-\beta) \Delta} e^{-\beta'\Delta} \leq e^{(\beta'-\beta) \varepsilon} \Theta_H(\beta').
\eea
Using modular covariance \eqref{modularc}, positive-definiteness of $\Theta_L$ and \eqref{main} we find
\bea
\left({ \beta\over 2\pi}\right)^c \Theta_L(\beta)&\geq& \left({ \beta\over 2\pi}\right)^c \Theta_L(\beta) -\Theta_L(\beta') =\\
& &\Theta_H(\beta')- \left({ \beta\over 2\pi}\right)^c \Theta_H(\beta) \geq  \left(1 -
 \left({ \beta\over 2\pi}\right)^c
e^{(\beta'-\beta) \varepsilon}\right)\Theta_H(\beta'). \nonumber
\eea
Using \eqref{main} one more time and assuming $f(\beta)<1$ we find
\bea
\Theta_L(\beta) {f(\beta)\over 1-f(\beta)}\geq \Theta_H(\beta),\qquad 
f(\beta):=\left({ \beta\over 2\pi}\right)^c
e^{(\beta'-\beta)\varepsilon}.
\eea
Finally we have
\bea
\Theta_L(\beta) \leq \Theta_\Lambda(\beta) \leq \Theta_L(\beta) {1\over 1-f(\beta)}.
\eea

To make sure $f(\beta)<1$ for all $\beta>2\pi$, $\varepsilon/c$ must be larger than 
\bea
{\ln(\beta/2\pi)\over \beta-\beta'} \label{vare}
\eea 
for all $\beta>2\pi$. The minimum of \eqref{vare} is achieved at $\beta=2\pi$ and is equal to $1/(4\pi)$. Hence 
\bea
\varepsilon={c\over 4\pi}+\epsilon
\eea
where $\epsilon$ is some positive constant, which should be chosen such that $\epsilon/c\rightarrow 0$. With this choice of $\varepsilon$ for any $\beta\geq 2\pi$,
\bea
f(\beta)\leq e^{(\beta'-\beta)\epsilon},
\eea
and we arrive at the analog of the HKS result for Narain theories
\bea
\label{NHKS}
\ln(\Theta_L(\beta)) \leq \ln(\Theta_\Lambda (\beta)) \leq \ln(\Theta_L(\beta))-\ln(1-e^{(\beta'-\beta)\epsilon}),\qquad \beta>2\pi.
\eea
With the conventional procedure of taking $\epsilon$ to zero, we recognize the logarithmic term in \eqref{NHKS} to be of subleading order $o(c)$.\footnote{Exactly as in the original paper \cite{hartman2014universal} this argument works as long as $(\beta-2\pi)$ is sufficiently large and $-\ln((\beta-2\pi)\epsilon)\ll c$.} 

Now we can introduce the notion of a sparse Narain theory, a theory for which 
\bea
\ln(\Theta_L(\beta)) 
\eea
is of order $O(1)$ in the $1/c$ expansion. In such a theory  $\Theta_\Lambda$ is given at leading order by (compare with \eqref{thetagrav})
\bea
\ln(\Theta_\Lambda)=\ln\left\{1+\left({2\pi\over \beta}\right)^c\right\}+o(c)
\eea
for all values of $\beta$, as follows from modular covariance.
In terms of the density of primaries the sparseness condition is  (compare with \eqref{sparseness})
\bea
\varrho(\Delta)\leq e^{2\pi \Delta},\qquad 0<\Delta \leq {c\over 4\pi}.
\eea
Sparseness  is necessary and sufficient for 
\bea
\ln(\Theta_\Lambda(\beta))=c\ln(2\pi/\beta)+O(1),\qquad \beta<2\pi,
\eea
and for the validity of the Cardy formula \eqref{Cl}\footnote{An analogous result was independently derived by T.~Hartman (private communication).}
\bea
\ln\varrho=c \ln(2\pi e \Delta/c)+O(1),\qquad \Delta\geq {c\over 2\pi}.
\eea

To summarize, the emergent picture is completely analogous to the analysis of \cite{hartman2014universal}: sparse theories are those for which the free energy matches the free energy of pure gravity ($U(1)$ gravity in our case) for all $\beta$ at leading order in $1/c$. For such theories the Cardy formula applies  already for $\Delta/c$ of order one.

While the averaged Narain theory is sparse, we do not know of any explicit example of an individual (non-averaged) large-$c$ Narain theory with this property.   This echoes the situation described in the introduction, that no explicit example of an HKS-sparse CFT with a weakly coupled gravity dual is known, although hypothetically such CFTs dominate the ensemble of all theories. In particular, none of the code CFTs introduced in section \ref{sec:codeCFTs} are sparse. Indeed, any code includes the trivial codeword $\c=0$ and hence the density of primary states at small $\Delta$ is given by
\bea
\ln \varrho=c\, \lambda(\alpha,0),
\eea
which violates the sparseness condition, as shown in  the inset of Fig.~\ref{fig:MO}.

\begin{figure}[t]
\begin{center}
\includegraphics[width=0.8\textwidth]{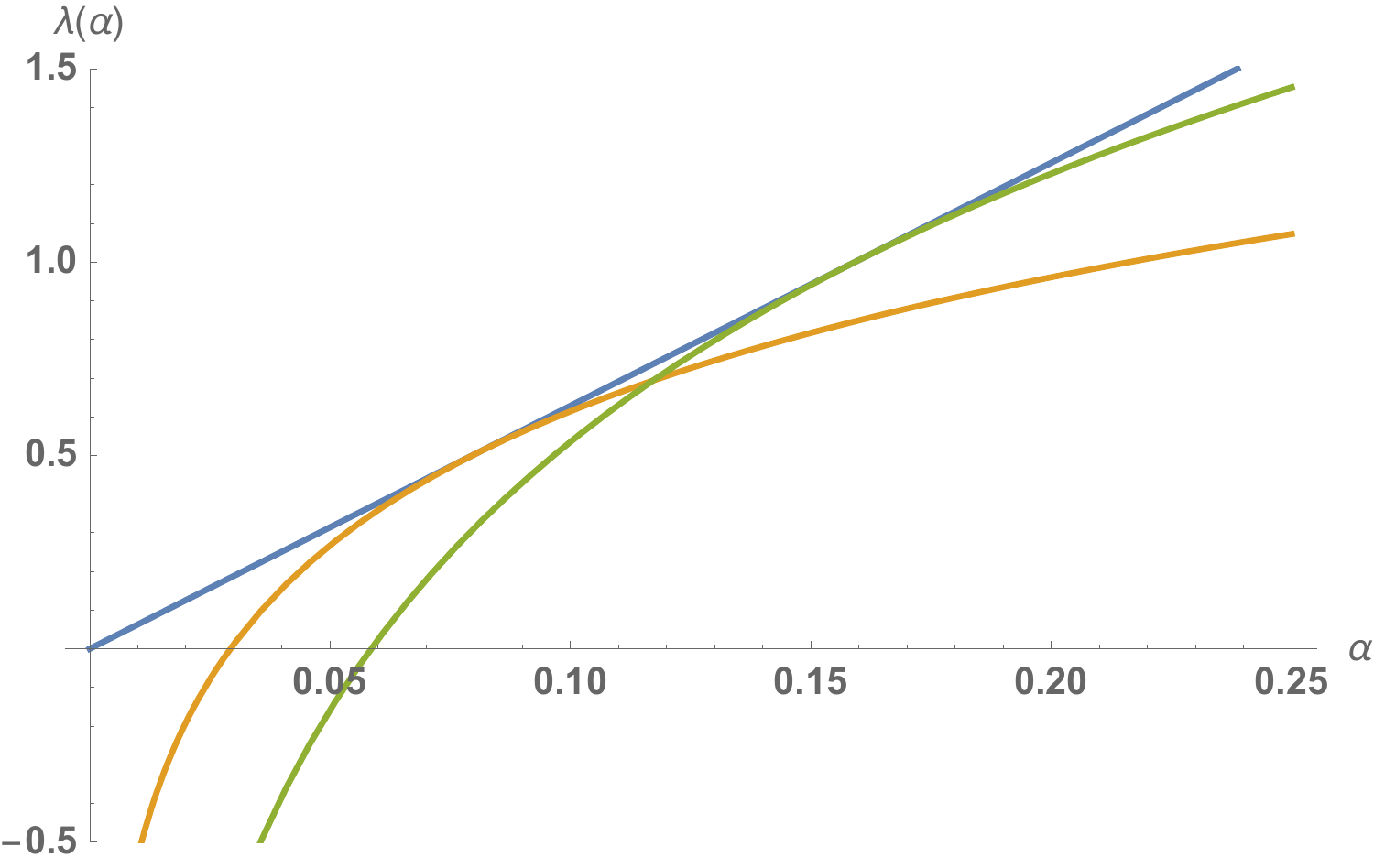}
\end{center}
\caption{Sparseness condition $2\pi \alpha$ (blue line) vs.~density of states of averaged chiral theory, $\lambda_{\it ch}(\alpha)$ (orange) and $\lambda_C(\alpha)$ (green).}
\label{fig:chiral}
\end{figure}

The HKS analysis also applies to chiral theories. Since the averaged chiral lattice CFT is obviously sparse, given a typical chiral theory (of which no explicit examples are known) we can construct a sparse Narain CFT as follows: every even self-dual lattice $\Uplambda \subset \R^{c}$ gives rise to a Narain lattice $\Lambda=\Uplambda_L\oplus \Uplambda_R \subset \R^{c,c}$. Its partition function is simply
\bea
Z={\Theta_\Uplambda(\tau)\Theta_\Uplambda(\bar\tau)\over |\eta|^{2c}}.
\eea
Then the spectral density at leading order can be obtained by the convolution of \eqref{densitychiral} with itself (because the variance is vanishingly small, as will be shown below)
\bea
N(\Delta)=1+2 {(2\pi \Delta)^{c/2}\over \Gamma(c/2+1)}+{(2\pi \Delta)^{c}\over \Gamma(c+1)}.
\eea
In terms of $\lambda$ there are two contributions, 
\bea
\lambda_{\it ch}(\alpha)={\ln(4\pi e \alpha)\over 2}
\eea
coming from the second term and 
$\lambda_C=\ln(2\pi e \alpha)$ coming from the last term. It is clear that $\lambda_{\it ch}$ satisfies the sparseness condition as is shown in Fig.~\ref{fig:chiral}.

As a side comment we notice that the averaged chiral CFT has spectral gap of $\Delta_1=c/(4\pi e)$, i.e.~a Narain CFT based on a random Euclidean even self-dual lattice would have this value of spectral gap. This agrees with the analogous result of \cite{afkhami2020free}. 

\section{Main hypothesis and its consistency}
\label{sec:analysis}
In thinking about the bulk description of conventional (Virasoro) large $c$ holographic theories, we usually take pure quasiclassical gravity to be the minimal ``core'' theory which can be combined with other states (potentially described by matter fields) in the bulk. We would like to probe the consistency of this picture for Narain theories, with $U(1)$ gravity playing the role of the minimal ``core.'' We emphasize that our discussion refers not to a hypothetical microscopic bulk description, but to effective IR behavior. Provided this picture is correct, we arrive at the following main hypothesis: the density of primary states of any Narain theory is bounded from below by the density of states of $U(1)$ gravity -- the Cardy formula \eqref{Cl},
\bea
\label{conjecture2}
{\ln\varrho\over c}\geq \lambda_C=\ln(2\pi e \alpha),\qquad \alpha=\Delta/c.
\eea
Since, by assumption, this inequality applies to all Narain theories and $\lambda_C$ is obtained by averaging over all 
Narain theories, the inequality \eqref{conjecture2} applies only in the strict $c \rightarrow \infty$ limit. The LHS of \eqref{conjecture2} should be understood as a limit taken with respect to any family of Narain theories defined for arbitrarily large $c$, provided the limit converges. This hypothesis immediately implies an asymptotic value of the maximal spectral gap  of primary states
\bea
\max \, \Delta_1={c\over 2\pi e}. \label{spectralgap}
\eea
Here we define $\Delta_1$ as the minimal value for which the density of states is exponentially large.  The conventional spectral gap, i.e.~the dimension of the lightest non-trivial primary, is less than or equal to $\Delta_1$ and therefore  satisfies $\Delta_1 \leq {c/ 2\pi e}$. As we will argue below, the distribution of the density $\varrho$ around the mean value $e^{\lambda_C}$ is vanishingly narrow in the large-$c$ limit, hence for nearly all large-$c$ theories $\Delta_1={c/2\pi e}$.

An immediate question is to assess the consistency of our hypothesis  \eqref{spectralgap} with the behavior of $\Delta_1$ for small and intermediate $c$ theories, coming from the numerical modular bootstrap. The results of the spinning modular bootstrap  for $c\leq 15$ \cite{afkhami2020free} suggest the behavior $\Delta_1\approx c/8+1/2$\,\footnote{Strictly speaking \cite{afkhami2020free} calculates the  bound on $\Delta_1$, but for $c\leq 8$ the authors also conjectured the maximal value of $\Delta_1$.},  seemingly in stark disagreement with \eqref{spectralgap}. It should be immediately noted this behavior is an artifact of small $c$. Indeed, $\Delta_1$ is bounded from above by the spinless modular bootstrap of \cite{afkhami2020high}, which uses numerical results for $c\leq 2000$ to estimate a stronger asymptotic bound of $c/9.869$. For smaller values of $c$, the bounds on $\Delta_1$ following from the spinless bootstrap are conservative; hence one can safely conclude that any asymptotic behavior satisfying $\Delta_1\leq c/9.869$ is consistent with the available numerics. 

Next, we would like to study the implications of our hypothesis for code theories. The inequality \eqref{conjecture} implies an upper bound on the binary distance $d_b$ of any real self-dual stabilizer code. The density of states of a code theory is dominated by the contribution of the zero codeword for small $\alpha$, but contributions of other codewords appear for $\alpha\geq d_b/(4c)$.
To make sure that $\lambda(\alpha)$ of a given code theory does not dip below $\lambda_C(\alpha)$, new primary states must appear at or below the value of $\alpha^*\approx
0.08993$ where the contribution of the zero codeword is equal to the Cardy formula (intersection of the blue and brown lines in Fig.~\ref{fig:GV})
\bea
\lambda(\alpha^*,0)=\lambda_C(\alpha^*).
\eea  
This implies a new asymptotic inequality 
\bea
{d_b\over c}\leq 4\alpha^*\approx 0.3597. \label{bound}
\eea
Since the conventional Hamming distance is bounded by the binary one, $d\leq d_b$, we find $d/c\leq 0.3597$ which is consistent with, close to, but weaker than the linear programming bound $d/c\leq 1/3$ \cite{rains1999quantum}. On the other hand, real self-dual stabilizer codes, via the Gray map discussed in \cite{Dymarsky:2020qom}, can be understood as binary isodual $[2n,n,d_b]$ codes. Therefore $d_b/c$ is bounded from above by the linear programming bound for formally self-dual binary codes, 
$d_b/c\leq (1-3^{-1/2})\approx 0.4226$ \cite{rains2003new}. Again, this is consistent with \eqref{bound}. Finally, isodual codes include even self-dual codes as a particular subclass, for which the bound is stronger $d_b/c\leq (1-5^{-1/4})\approx 0.3313$ \cite{rains2003new}. In other words the new bound \eqref{bound} is consistent with other bounds found in the literature, which serves as an indirect consistency check of the main conjecture. 

\begin{figure}[t]
\begin{center}
\includegraphics[width=0.8\textwidth]{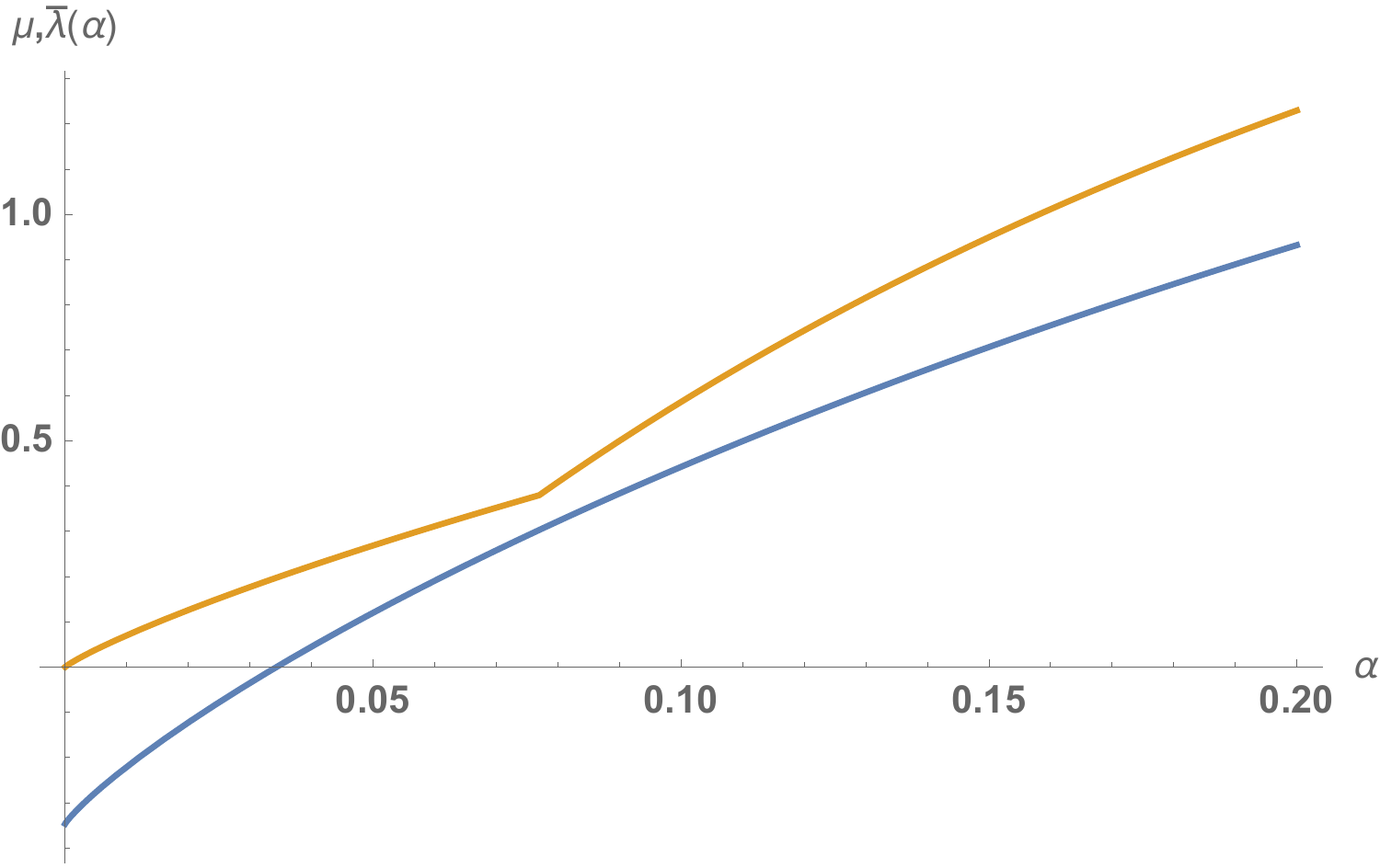}
\end{center}
\caption{Leading (exponential) contribution to  the square root of the variance of the density of states (blue line) vs.~mean density of states (orange line).}
\label{fig:mu}
\end{figure}
Finally, in order for the main conjecture to hold, a necessary condition is for the variance of $\varrho$ around its mean value to be vanishingly small in the large $c$ limit, $\overline{\delta \varrho^2}(\Delta)/(\overline{\varrho}(\Delta))^2 \rightarrow 0$, where 
\bea
\delta \varrho(\Delta):=\varrho(\Delta)-\overline{\varrho}(\Delta),
\eea
and the bar stands for averaging over the ensemble of Narain theories. In lieu of this calculation\footnote{Certain contributions to variance in the Narain case were calculated in \cite{maloney2020averaging,cotler2020ads1}. They are $e^{-O(c)}$ suppressed, which is consistent with our expectations.} we make use of the similarity between the problems of maximizing the CFT spectral gap for both Narain and chiral theories, finding sphere packings of maximal density, and finding optimal classical and quantum codes (all three problems can be understood as finding a lattice from a given class with the largest possible shortest vector), and first perform an average over the space of code CFTs. The underlying assumption here is that in all these cases the qualitative behavior is the same.  The average spectral density of code theories was found in section \ref{sec:codeCFTs},
\bea
\overline{\varrho}(\Delta)=e^{c {\overline \lambda}(\alpha)}=e^{c {\overline \lambda}_1(\alpha)}+e^{c{\overline \lambda}_2(\alpha)},\qquad \alpha=\Delta/c,
\eea
in the large-$c$ limit. In Appendix \ref{app:B} we calculate the variance of the averaged refined enumerator polynomial, see eq.~\eqref{variance}, which gives
\bea
&&\overline{\delta \Theta(\beta_1) \delta \Theta(\beta_2)}=\left({\theta_3(q_1^2)\theta_3(q_2^2)+\theta_3(q_1^2)\theta_3(q_2^2) \over \sqrt{2}}\right)^{2c},\\
&&\delta \Theta(\beta)= \Theta(\beta) -\overline{\Theta}(\beta),\quad q_1=e^{-\beta_1},\, \, q_2=e^{-\beta_2}. \nonumber
\eea
To calculate the connected two-point correlation function we need to perform an inverse Laplace transformation  
\bea
\overline{\delta \varrho(\Delta_1)\delta \varrho(\Delta_2)}= e^{c f+\beta_1 \Delta_1+\beta_2 \Delta_2},
\eea
where $\beta_i$ is fixed by 
\bea
{\partial f\over \partial \beta_i}=-\alpha_i,\qquad f=2\ln(\theta_3(q_1^2)\theta_3(q_2^2)+\theta_3(q_1^2)\theta_3(q_2^2))-\ln(2).
\eea
For equal $\Delta_1=\Delta_2$ we plot $\mu(\alpha)$ where
\bea
\overline{\delta \varrho^2}(\alpha c)=e^{2c \mu (\alpha)}
\eea
in Fig.~\ref{fig:mu}, superimposed with $\overline{\lambda}_{1,2}$. It is clear that for code theories the variance is exponentially suppressed for all $\Delta$
\bea
{\overline{\delta \varrho^2}(\Delta)\over (\overline{\varrho}(\Delta))^2}= e^{2c(\mu-\overline{\lambda})}\propto e^{-O(c)},
\eea
confirming the overall picture.

Next we perform a similar calculation for the ensemble of chiral theories. Our starting point is the two-point correlator  
\bea
\overline{\Theta(\tau_1)\Theta(\tau_2)}=\sum_{n,m} c_{nm}\, q_1^n q_2^m.
\eea
Here the integers $n,m$ should be interpreted as dimensions and
\bea
\overline{\varrho(n)\varrho(m)}=c_{nm}.
\eea
An explicit expression for $c_{nm}$ was obtained in \cite{waldspurger1978generation,eichler1985theory}
\bea
c_{nm}={2\over \zeta(1-k)}\sum_{d|(n,m)} d^{k-1} \sum_{r^2\leq {4nm\over d^2}} H\left(k-1,{4nm\over d^2}-r^2\right). \label{cnm}
\eea
After certain manipulations we arrive at the following representation for the functions $H$, convenient for taking the large-$k$ limit,
\bea
\label{h}
H(k-1,N)&=&\sum_{d|N} h(k-1,N/d^2),\\
h(r,N)&=&\left\{\begin{array}{cc}
\tilde{h}, & (-1)^r N\, {\rm mod}\,  4=0,1, \\
0, & (-1)^r N\, {\rm mod}\,  4=2,3. 
\end{array}\right.\qquad \tilde{h}={2\, \Gamma(r) N^{-1/2}\over (2\pi)^r \zeta(1-2r)} \sum_{l=1}^N \left({N|l}\right)\zeta(r,l/N). \nonumber
\eea
Here $\left({N| k}\right)$ is Kronecker symbol. We have checked this representation by reproducing the numerical values of $H(k-1,N)$ given in the appendix of \cite{cohen1975sums}.

In the limit $r=k-1\gg 1$ Hurwitz zeta function $\zeta(r,k/N)$ drastically simplifies 
\bea
\zeta(r,l/N) \approx \left({N\over l}\right)^{r}.
\eea
From here it follows that the $l=1$ term in the sum contributes at leading order, while the contributions with $l\geq 2$ are exponentially suppressed, $e^{-O(r)}$,
\bea
\tilde{h}(k-1,N)\approx {(2\pi)^{k-1} \Gamma(k-1) N^{k-3/2}\over \Gamma(2k-2)},
\eea
where we have approximated $\zeta(3-2k)\approx 1$ with exponential precision.
Our next step is to sum  over $r$ in \eqref{cnm}, which can be substituted by integration with exponential precision in the regime of interest, $k\rightarrow 
\infty$ for fixed $n/k,m/k$, as follows from the Euler--Maclaurin formula (see Appendix \ref{sec:C}),
\bea
\sum_{r^2\leq N} H(k-1,N-r^2) \approx \int_{-D^{1/2}}^{D^{1/2}} dr (N-r^2)^{k-3/2} {(2\pi)^{k-1} \Gamma(k-1)\over \Gamma(2k-2)}=\frac{2^{2-k} \pi ^k}{\Gamma (k)} N^{k-1}. \nonumber
\eea
Finally we notice that in both \eqref{cnm} and \eqref{h} only $d=1$ contributes at leading order, while the contributions of other terms are exponentially suppressed, as $1/d^k$. Thus we find
\bea
c_{nm}={(2\pi)^{2k} \over \Gamma(k)^2} \left({nm}\right)^{k-1} \left(1+e^{-O(k)}\right).
\eea
Comparing with \eqref{densitychiral} confirms that the connected two point function is exponentially suppressed 
\bea
\overline{\rho(\Delta_1)\rho(\Delta_2)}=\overline{\varrho(\Delta_1)}\cdot \overline{\varrho(\Delta_2)} \left(1+e^{-O(k)}\right).
\eea

Certainly, a vanishing variance, provided that an analogous result holds for the ensemble of all Narain theories, does not guarantee the validity of \eqref{conjecture2}. For that, a much stronger property is required: that the distribution of $\varrho$ around the mean has no fat tails, which would allow substantial deviations from the mean even as $c\rightarrow \infty$. Studying the higher moments of $\delta \varrho$ could provide additional consistency checks, but not a proof, unless one gains theoretical control over arbitrarily high moments, similarly to  \cite{engelhardt2020free}.

\section{Discussion}
\label{sec:discussion}
In this paper we have discussed a scenario in which the holographic dual of a given  Narian CFT in the infinite $c$ limit is effectively described as ``$U(1)$ gravity'' of \cite{afkhami2020free,maloney2020averaging} together with some additional states (fields). This simple picture, if correct,  would imply that pure $U(1)$ gravity 
has the lowest possible density of states, leading to our main conjecture \eqref{conjecture2}. We have tested this hypothesis by noting that it is consistent with available numerics; it leads to new bounds on quantum codes which are consistent with previously known ones; and we have argued the variance of the density $\varrho(\Delta)$ is vanishingly small, as is required by \eqref{conjecture2}, by considering  two related ensembles. Our analysis amounts to a series of indirect consistency checks and thus can not directly establish the validity of \eqref{conjecture2}. 

More direct evidence could come from the numerical conformal bootstrap in the form of a bound on the maximal spectral gap \eqref{spectralgap}. One would need to extend the numerical analysis of \cite{afkhami2020free} to higher values of  $c$ until the asymptotic slope emerges.  While the result could turn out to be overly conservative (as is currently believed to be the case with the bound on the spectral gap of Virasoro theories), the problem of finding the maximal spectral gap can be formulated as a quadratic optimization problem, which makes it plausible that this question could be studied numerically.  

Another potential way to confirm the scenario in question could come from the bulk microscopic description, e.g.~in terms of Chern-Simons gauge fields with additional matter, dual to a given Narain theory.
The microscopic description by itself will likely be insufficient: it will presumably depend on the metric $G$ and $B$-field parameterizing Narain moduli space,  but evaluating, for example, the spectral gap $\Delta_1(G,B)$ for given  $G$ and $B$ would likely be as complicated on the bulk side as it is in the boundary theory. A crucial new ingredient necessary on the bulk side would be an analog of the Bekenstein-Hawking entropy formula for gauge fields \cite{takayanagi2020chern,zhao2020symmetry}, obtained potentially via embedding gauge theory into bona fide gravity in $AdS_3$ \cite{perez2020gravitational}.
Presumably, if this formula were applied to the ``BTZ'' configuration discussed in section \ref{sec:HKS} it would yield the Cardy formula \eqref{Cardy}, mirroring the conventional analysis for Virasoro theories and establishing the validity of \eqref{conjecture2}. 

The underlying assumption we relied on in section \ref{sec:analysis} is that the spaces of Narain theories, of chiral CFTs  defined by even self-dual Euclidean lattices, of Narain theories associated with quantum codes, and of lattice sphere packings (arbitrary lattices with the unit cell volume) are not only closely related (as discussed  in \cite{hartman2019sphere,Dymarsky:2020bps}), but in fact share crucial features. Thus, we expect that the vanishing variance of the level densities of code and chiral theories will extend to the ensemble of all Narain theories. By extending this reasoning to lattice sphere packings we arrive at the well-known lore that a random sphere packing is as effective as the averaged one \cite{elkies2000lattices}. Furthermore, in  asymptotically large dimensions a random sphere packing is conceivably the densest one  \cite{elkies2000lattices}, mirroring our conjecture for the maximal spectral gap \eqref{spectralgap}. Should our scenario get confirmed, possibly as outlined above, that would in turn solidify the expectation for the sphere packing problem, providing non-trivial feedback from quantum gravity to discrete geometry. 

Confirming the scenario of our paper would also support the expectation that pure gravity in $AdS_3$ is dual not to a particular theory but to an appropriately defined ensemble. An important difference between the original analysis of Hartman, Keller and Stoica and the analysis of section \ref{sec:HKS} is that in the Virasoro case the upper limit of the range of the sparseness condition $c/12$ \eqref{sparseness} coincides with the lowest value of $\Delta$ for which Cardy formula yields positive level density. In other words, Virasoro theories with maximal spectral gap are automatically sparse. This is not the case for Narain theories, as the upper limit of the sparseness condition $c/(4\pi)$ exceeds the hypothetical maximal spectral gap $c/(2\pi e)$. In other words, there are likely  non-sparse Narain theories with maximal spectral gap. We interpret this as an indication that there are different classes of Narain theories, not only those satisfying the sparseness condition, which can be described holographically using quasi-classical fields in the bulk. On the contrary, in the Virasoro case there seems to be only one class of sparse theories with a weakly-coupled bulk description. \\

%\acknowledgements
We thank  O.~Aharony, T.~Hartman, A.~Levin, A.~Maloney, and E.~Perlmutter for discussions. This paper benefited from the talks and discussions of the Simons Center for Geometry and Physics workshop Sphere Packing and the Conformal Bootstrap. A.D.~is grateful to Weizmann Institute of Science for hospitality and acknowledges   sabbatical support of the Schwartz/Reisman Institute for Theoretical Physics,
and support by the NSF under  grant PHY-2013812.

\appendix
\section{Density of states of code theories}
\label{app:A}
In this section we derive the key result of Mazo and Odlyzko \cite{mazo1990lattice} (\ref{MO1}, \ref{MO2}) at a physicist's level of rigor. Starting from the definition of the Jacobi theta functions 
\bea
\theta_3(q)=\sum_n q^{n^2/2},\qquad \theta_2(q)=\sum_n q^{(n+1/2)^2/2},\qquad q=e^{2\pi i \tau},
\eea
we easily calculate the contribution of a given codeword $\c\in \C$ to the lattice theta-function provided $\C$ is understood as a binary $[2c,c,d_b]$ code,
\bea
\Theta(\c)=\theta_3(q^2)^{2c-w(\c)} \theta_2(q^2)^{w(\c)}. 
\eea
This is of course the same as \eqref{Z} after the substitution $\tau=-\bar \tau= i\beta/(2\pi)$. For convenience we introduce $p=w(\c)/c$ and represent $Z(\c)$ as 
\bea
\Theta(\c)=\sum_\Delta  \varrho(\Delta) e^{-\beta \Delta}.
\eea
Using the saddle point approximation we immediately find
\bea
{\Delta(\beta)\over c}=-{\partial \ln \Theta\over c\, \partial \beta}=-{\partial \over \partial \beta}\left( (2-p)\ln \theta_3(q^2)+p \ln \theta_2(q^2)\right),\qquad q=e^{-\beta}.
\eea
Upon redefinition $q^2\rightarrow q$ and $\Delta/c=\alpha$ this becomes \eqref{MO2}.
The density of states follows from here
\bea
{\ln \varrho\over c}={\ln \Theta(\c)+\beta \Delta\over c},
\eea
which gives \eqref{MO1}.

As an illustration of consistency we calculate $\overline{\lambda}_2$ -- the contribution of the term $(1+t)^{2c}/2^c$ from \eqref{GV} toward the full density of primaries, 
\bea
N(\Delta) \approx {1\over 2^c}\sum_{k=0}^{2c} C_{2c}^k\, e^{c \lambda(\alpha,k/c)}.
\eea
The leading (saddle point) contribution is given by the value of $k$ such that 
$0\leq p^*=k/c\leq 2$ maximizes
\bea
\lambda(\alpha,p)+2H(p/2)-\ln(2),\qquad H(p)=-p \ln(p) -(1-p)\ln(1-p),
\eea
and
\bea
\label{2}
{\overline \lambda}_2=\lambda(\alpha,p^*)+2H(p^*/2)-\ln(2).
\eea
The condition 
\bea
{\partial\over \partial p}\left(\lambda(\alpha,p)+2H(p/2)\right)=0
\eea
can be solved analytically
\bea
p^*={2\theta_2(q^2)\over \theta_3(q^2)+\theta_2(q^2)}.
\eea
After substituting this back into \eqref{2} we recover \eqref{lambda2}.

\section{Averaging over code theories}
\label{app:B}
In this section we denote $c$ by $n$, which means we are considering self-dual $[[n,0,d]]$ quantum codes. 
\subsection{Averaged enumerator polynomial}
We consider the ensemble of all $B$-form codes introduced in \cite{Dymarsky:2020qom}. The codes are parametrized by binary symmetric $n\times n$ matrices $B$ with zeroes on the diagonal. The space of all such matrices will be denoted as $M_n$.
For a given $B$ the $2^n$ codewords are given by $\c(B,x):=(x_1,\dots,x_n,B\vec x)$ where $\vec{x}$ is an arbitrary binary vector of length $n$. Hence, for pure imaginary $\tau$ the
averaged refined enumerator polynomial is given by 
\bea
\overline{W}(1,t^2,t)=\sum_{x \in \Z_2^n} {1\over 2^{n(n-1)/2}} \sum_{B\in M_n} t^{w(B,x)},\\
w(\c(B,x))=w(B,x)=\vec{1}\cdot \vec{x}+\vec{1}\cdot (B \vec{x})_2,
\eea
where $(B \vec{x})_2$ is the binary vector congruent to $(B \vec{x}) \ {\rm mod} \ 2$ .
Without loss of generality we can assume that $\vec{x}$ is ordered, $\vec{x}=(\underbrace{1,\dots,1}_a,0,\dots,0)$ and $\vec{1}\cdot \vec{x}=a$, for $0\leq a \leq n$. We would like to calculate 
\bea
Q_a(t)={1\over 2^{n(n-1)/2}} \sum_{B\in M_n} t^{\vec{1}\cdot (B \vec{x})_2},
\eea Obviously $Q_a(t)=(1+t)^{n-a}/2^{n-a}P_a(t)$ where
\bea
P_a(t)={1\over 2^{a(a-1)/2}} \sum_{B \in M_a} t^{\vec{1}\cdot (B \vec{1})_2},
\eea
unless $a=0$ in which case $Q_0(t)=1$.
We can calculate $P_a$ iteratively by adding one more row and column to matrix $B$. Assuming a particular configuration of $B$ would yield a term $t^k$, adding an additional row and column to $B$ would turn it into $t^i ((1+t)+(-1)^i(1-t))/2$ if $k$ were even or $t^i ((1+t)-(-1)^i(1-t))/2$ if k were odd. (Here of course we need to sum over all possible $i$ with the weight $C_a^i/2^a$.) In other words 
\bea
t^k \rightarrow {1\over 2^a}\sum_{i=0}^a C_a^i t^i ((1+t)+(-1)^{i+k}(1-t))/2= {(1+t)^{a+1}+(-1)^k(1-t)^{a+1}\over 2^{a+1}}.
\eea
We therefore arrive at the following iterative relation
\bea
P_{a+1}(t)= {P_a(1)(1+t)^{a+1}+P_a(-1)(1-t)^{a+1}\over 2^{a+1}}.
\eea
From here and $P_0(t)=1$ we find
\bea
P_a(t)\rightarrow {(1+t)^{a}+(1-t)^{a}\over 2^{a}}-\delta_{a,0}. \label{P}
\eea

Now we are ready to calculate 
\bea
\overline{W}(1,t^2,t)=1+\sum_{a=1}^n C_n^a\, t^a\, Q_a(t) =\\ 
1+{(1+t)^{2n}\over 2^n}+{(1+2t-t^2)^{n}\over 2^n}-2{(1+t)^n\over 2^n},
\eea
matching the general result of \cite{Dymarsky:2020qom}.

\subsection{Averaged square of enumerator polynomial}
Now we would like to find 
\bea
\overline{W(1,t_1^2,t_1)W(1,t_2^2,t_2)}=\sum_{x,y \in \Z_2^n} {1\over 2^{n(n-1)/2}} \sum_{B\in M_n} t_1^{w(B,x)} t_2^{w(B,y)}.
\eea
We can first assume that $\vec{x}=(\underbrace{1,\dots,1}_{m},\underbrace{1,\dots,1}_{k},\underbrace{0,\dots,0}_{l})$ and $\vec{y}=(\underbrace{1,\dots,1}_{m},\underbrace{0,\dots,0}_{k},\underbrace{1,\dots,1}_{l})$, while the matrix $B$ is $n\times n$ and $n=m+k+l$. We want to calculate 
\bea
P_{m,k,l}(t_1, \tilde{t}_1,\hat{t}_1,t_2,\tilde{t}_2,\hat{t}_2)={1\over 2^{n(n-1)/2}}\sum_{B\in M_n} t_1^{\vec{e}_m\cdot (B x)_2} \tilde{t}_1^{\vec{e}_k\cdot (B x)_2} \hat{t}_1^{\vec{e}_l\cdot (B x)_2} t_2^{\vec{e}_m\cdot (B y)_2} \tilde{t}_2^{\vec{e}_k\cdot (B y)_2} \hat{t}_2^{\vec{e}_l\cdot (B y)_2} \nonumber
\eea
where $\vec{e}_m=(\underbrace{1,\dots,1}_{m},\underbrace{0,\dots,0}_{k},\underbrace{0,\dots,0}_{l})$, 
$\vec{e}_k=(\underbrace{0,\dots,0}_{m},\underbrace{1,\dots,1}_{k},\underbrace{0,\dots,0}_{l})$ and finally $\vec{e}_l=(\underbrace{0,\dots,0}_{m},\underbrace{0,\dots,0}_{k},\underbrace{1,\dots,1}_{l})$ such that $x=e_m+e_k$ and $y=e_m+e_l$.

We can find $P_{m,k,l}$ iteratively by increasing $l$ by one (adding one additional row and column to $B$). Any monomial $t_1^a\, \tilde{t}_1^{\tilde{a}}\, \hat{t}_1^{\hat{a}} \,
t_2^b\, \tilde{t}_2^{\tilde{b}}\, \hat{t}_2^{\hat{b}}$ will turn into 
\bea
\nonumber
{1\over 2^n}\sum_{i=0}^m \sum_{\tilde{i}=0}^k \sum_{\hat{i}=0}^l C_m^i C_k^{\tilde{i}} C_l^{\hat i}t_1^a \tilde{t}_1^{\tilde{a}} \hat{t}_1^{\hat{a}}   t_2^i \tilde{t}_2^{\tilde{i}} \hat{t}_2^{\hat{i}}  \times \\ {(1+\hat{t}_1)+(-1)^{b +i + \tilde{b}+\tilde{i}\ {\rm mod}\  2}(1-\hat{t}_1)\over 2}
\times {(1+\hat{t}_2)+(-1)^{b +i + \hat{b}+\hat{i}\ {\rm mod}\  2}(1-\hat{t}_2)\over 2}.
\eea
From here we find
\bea
\nonumber
P_{m,k,l+1}={1\over 2^3}
\sum_{\epsilon=\pm 1}
\sum_{\nu=\pm 1}
\sum_{\mu=\pm 1}
\sum_{r=0}^1 \sum_{\tilde{r}=0}^1  \sum_{\hat{r}=0}^1 \epsilon^r \nu^{\tilde r} \mu^{\hat r} P_{m,k,l}(t_1,\tilde{t}_1,\hat{t}_1,(-1)^r,(-1)^{\tilde{r}},(-1)^{\hat r})  \times \\
\nonumber
{1\over 2^n}\sum_{i=0}^m \sum_{\tilde{i}=0}^k \sum_{\hat{i}=0}^l C_m^i C_k^{\tilde{i}} C_l^{\hat i}  t_2^i \tilde{t}_2^{\tilde{i}} \hat{t}_2^{\hat{i}}  \times \\ {(1+\hat{t}_1)+(-1)^{i + \tilde{i}}\epsilon \nu (1-\hat{t}_1)\over 2}
\times {(1+\hat{t}_2)+(-1)^{i  +\hat{i}} \epsilon \mu(1-\hat{t}_2)\over 2}=
\nonumber
\\ 
\nonumber
{1\over 2^{m+k+l+2}} P_{m,k,l}(t_1,\tilde{t}_1,\hat{t}_1,1,1,1) (1+t_2)^m (1+\tilde{t}_2)^k (1+\hat{t}_2)^{l+1} (1+\hat{t}_1)+\ \ \\ \nonumber
{1\over 2^{m+k+l+2}} P_{m,k,l}(t_1,\tilde{t}_1,\hat{t}_1,-1,-1,1) (1-t_2)^m (1-\tilde{t}_2)^k (1+\hat{t}_2)^{l+1} (1-\hat{t}_1)+\\ \nonumber
{1\over 2^{m+k+l+2}} P_{m,k,l}(t_1,\tilde{t}_1,\hat{t}_1,-1,1,-1) (1-t_2)^m (1+\tilde{t}_2)^k (1-\hat{t}_2)^{l+1} (1+\hat{t}_1)+\\ \nonumber
{1\over 2^{m+k+l+2}} P_{m,k,l}(t_1,\tilde{t}_1,\hat{t}_1,1,-1,-1) (1+t_2)^m (1-\tilde{t}_2)^k (1-\hat{t}_2)^{l+1} (1-\hat{t}_1).\ \ \ \ 
\eea
The following expression is a solution
\bea
&&P_{m,k,l}(t_1,\tilde{t}_1,\hat{t}_1,t_2,\tilde{t}_2,\hat{t}_2)=\\
&&{1\over 2^{2(m+k+l)}}\sum_{\epsilon=\pm 1}
\sum_{\nu=\pm 1} \sum_{\mu=\pm 1} 
(1+\epsilon t_1)^m
(1+\epsilon \mu \tilde{t}_1)^k
(1+\mu \hat{t}_1)^l
(1+\nu t_2)^m
(1+\mu \tilde{t}_2)^k
(1+\mu \nu \hat{t}_2)^l. \nonumber
\eea
from where follows
\bea
\label{Pmkl}
&&P_{m,k,l}(t_1,t_1,t_1,t_2,t_2,t_2)=\\
&&{1\over 2^{2(m+k+l)}}\sum_{\epsilon=\pm 1}
\sum_{\nu=\pm 1} \sum_{\mu=\pm 1} 
(1+\epsilon t_1)^m
(1+\epsilon \mu t_1)^k
(1+\mu t_1)^l
(1+\nu t_2)^m
(1+\mu t_2)^k
(1+\mu \nu t_2)^l. \nonumber
\eea
This expression is correct only if $k+l>0$. Otherwise 
\bea
P_{m,0,0}(t_1,t_2)=P_m(t_1 t_2)={(1+t_1 t_2)^m+(1-t_1 t_2)^m\over 2^m}-\delta_{m,0}
\eea
as follows from \eqref{P}.

Now, if $n<m+k+l$ the expression for 
${1\over 2^{n(n-1)/2}} \sum_{B} t_1^{w(B,x)} t_2^{w(B,y)}$ where $x,y$ are characterized by $m,k,l$ is given by 
\bea
Q_{m,k,l}={1\over 2^{2(n-m-k-l)}}t_1^{m+k}(1+t_1)^{n-m-k-l} t_2^{m+l} (1+t_2)^{n-m-k-l} P_{m,k,l}(t_1,t_2), 
\eea
unless $k=l=0$ in which case we get 
\bea
Q_{0,0,l}=(t_1 t_2)^m (1+t_1 t_2)^{n-m} P_{m,0,0}(t_1,t_2), \label{m}
\eea
when $m>0$
and similarly for $m=l=0$, $k>0$ 
\bea
Q_{0,k,0}(t_1)^k (1+t_1)^{n-k} P_{m}(t_1), \label{k}
\eea
and  $m=k=0$, $l>0$ with the substitution $t_1 \leftrightarrow t_2$.
Finally we need to sum over $\vec{x}$ and $\vec{y}$, which is equivalent to summing over $m,k,l$ with the weight $\sum_{n\geq m+k+l\geq  0} {n!\over m! k! l! (n-m-k-l)!}$,
\bea
\nonumber
\sum_{n\geq m+k+l\geq  0} {n!\over m! k! l! (n-m-k-l)!}  t_1^{m+k}(1+t_1)^{n-m-k-l} t_2^{m+l} (1+t_2)^{n-m-k-l} P_{m,k,l}(t_1,t_2).
\eea
This is a naive expression because we included the full range of $m,k,l$. Instead we should subtract 
\bea
\label{pm}
{n!\over m! (n-m)!}t_1^{m}(1+t_1)^{n-m} t_2^{m} (1+t_2)^{n-m} P_{m,k,l}(t_1,t_2)\delta_{k+l=0}+\\  
\label{pk}
{n!\over k! (n-k)!} t_1^{k}(1+t_1)^{n-k}  (1+t_2)^{n-k} P_{m,k,l}(t_1,t_2)\delta_{m+l=0}+\\
{n!\over l!(n-l)!}
(1+t_1)^{n-l} t_2^l  (1+t_2)^{n-l} P_{m,k,l}(t_1,t_2)\delta_{m+k=0}
\label{pl}
\eea
and add terms \eqref{m}, \eqref{k} and similarly for $m=k=0$,
\bea
\overline{W}(1,t_1^2 t_2^2,t_1 t_2)+\overline{W}(1,t_1^2,t_1)+\overline{W}(1,t_2^2,t_2). \label{WWW}
\eea
Here we need to be careful: the first term in \eqref{WWW} stands for the sum over $m$ while $k=l=0$, the second term for the sum over $k$ with $m=l=0$, and the third for the sum over $l$ while $m=k=0$. We have therefore calculated the contribution of $m=k=l=0$ three times instead of once. When $m=k=l=0$ this means that $\vec{x}=\vec{y}=0$ and hence the contribution is simply $1$. We therefore must subtract $-2$.

We made a similar mistake above when subtracting \eqref{pm}, \eqref{pk}, \eqref{pl} -- we counted the contribution of $m=k=l$ three times instead of one. We therefore should subtract it twice with a minus sign 
\bea
2 \sum_{\epsilon=\pm 1}
\sum_{\nu=\pm 1} \sum_{\mu=\pm 1}  {(1+t_1)^n (1+t_2)^n\over 4^n} =16  {(1+t_1)^n (1+t_2)^n\over 4^n}.
\eea

Finally we get 
\bea
\nonumber
\overline{W(1,t_1^2,t_1)W(1,t_2^2,t_2)}=\overline{W}(1,t_1^2 t_2^2,t_1 t_2)+\overline{W}(1,t_1^2,t_1)+\overline{W}(1, t_2^2,t_2)-2+16  {(1+t_1)^n (1+t_2)^n\over 4^n}+\\  \nonumber
\sum_{\epsilon,\nu,\mu=\pm 1} {((1 + t_1) (1 + t_2) + t_1 t_2 (1 + \epsilon t_1) (1 + \nu t_2) +    t_1 (1 + \epsilon \mu t_1) (1 + \mu t_2) +    t_2 (1 + \mu t_1) (1 + \mu \nu t_2))^n\over 4^n}\\  \nonumber
-2\sum_{\epsilon,\nu=\pm 1}  {((1 + t_1) (1 + t_2) + t_1 t_2 (1 + \epsilon t_1) (1 + \nu t_2))^n\over 4^n}  \\ \nonumber
-2\sum_{\epsilon,\nu=\pm 1}  {((1 + t_1) (1 + t_2) + t_1 (1 + \epsilon t_1) (1 + \nu t_2))^n\over 4^n} \\ \nonumber
-2\sum_{\epsilon,\nu=\pm 1}  {((1 + t_1) (1 + t_2) + t_2 (1 + \epsilon t_1) (1 + \nu t_2))^n\over 4^n}.
\eea
From here follows that 
variance, at leading order, is given by
\bea
\overline{W(1,t_1^2,t_1)W(1,t_2^2,t_2)}-\overline{W(1,t_1^2,t_1)}\cdot \overline{W(1,t_2^2,t_2)}=
\left({(1+t_1 t_2)^2\over 2}\right)^n, \label{variance}
\eea
where we dropped all exponentially-suppressed terms. 

\section{Approximating sums by  integrals} %an?
\label{sec:C}
We consider the sum over $r$ from the equation \eqref{cnm}, which in the appropriate limit  $k\gg 1$ becomes 
\bea
I=\sum_{r=-p}^p \left(N-r^2\right)^{k-3/2},
\eea
where $p$ is the largest integer number such that $p^2\leq N$. We are interested in the  relative error, which would arise if we substitute summation for integration. It is convenient to divide the integral by a power of $N$, redefine $k\rightarrow k+3/2$, and change the lower bound of $r$ by one (this introduces an error of order  $p^{-k}\sim e^{-O(k)}$)
\bea
\tilde{I}=\sum_{r=-p+1}^p f(r),\qquad  f(x)=\left(1-{x^2\over N}\right)^{k}. \label{I}
\eea
Furthermore we assume that $N$ scales as $k^2$, i.e.~$p$ is of order $k$. 
The difference between the sum \eqref{I} and the integral (changing the limits of integration from $N^{1/2}$ to $p$ also introduces an error of order $e^{-O(k)}$)
\bea
\tilde{J}=\int_{-p}^{p} dx f(x)
\eea
is given by the Euler-Maclaurin formula
\bea
\tilde{I}-\tilde{J}={f(p)-f(-p)\over 2}-2\sum_{l=1}^{a} {\zeta(2l)\over (2\pi i)^{2l}}\left(f^{2l-1}(p)-f^{2l-1}(-p)\right)+R_{2a}. \label{EM}
\eea
Our goal is to show that all terms on the RHS of \eqref{EM} are exponentially suppressed. We start with $f(p)=(1-p^2/N)^k$. Since $p$ is the largest integer such that $p^2<N$, we can estimate $f(p)$ to be of order $p^{-k}$. Next, $f'(p)$ is of order $2kp/N (1/p)^{k-1}\propto 2 k/p^k$. Differentiating iteratively we find at leading order $f^{(l)}(p)\propto (2k)^l/p^k$, with each term contributing more to the sum. The contribution of the last (largest) term is of order $k^{2a}/p^k$ (all numerical coefficients are neglected).  Finally, the term $R_{2a}$ can be bounded by 
\bea
|R_{2a}|\leq {2 \zeta(2a)\over (2\pi)^{2a}}\int_{-p}^p dx \left|f^{(2a)}(x)\right|.
\eea
To estimate it, we calculate maximum value of $\left|f^{(2a)}(x)\right|$, which occurs at $x=0$, 
\bea
f^{(2a)}(0)={(-1)^a2^{2a-1}\over p^{2a}}{\Gamma(a+1/2)\Gamma(k+1)\over \Gamma(1/2)\Gamma(k+1-a)}.
\eea
After dropping all numerical coefficients we find at leading order $(ak/p^2)^a$. We will choose $a$ to make the discrepancy between $\tilde{I}$ and $\tilde{J}$ 
\bea
{k^{2a}\over p^k} +{(ak)^a\over p^{2a}}. \label{discrepancy}
\eea
sufficiently small.
Taking into account that $p$ scales as $k$, such that $\alpha=p/k$ is fixed 
we can take $a$ to be $a=\varepsilon k$ where $\varepsilon$ is chosen such that $\varepsilon/\alpha^2$ would be (much) smaller than one. Then in the limit $k \rightarrow \infty$ the first term in  \eqref{discrepancy} would scale as $e^{-k(1-2\varepsilon)\ln(k)}$ and the second term would scale as $e^{-k\, 2\varepsilon \ln(\alpha^2/\varepsilon)}$.
In other words  the discrepancy would be exponentially smaller than $\tilde{J}\sim p/k^{1/2}$. This estimate is likely to be non-optimal and can be improved. 

\bibliographystyle{JHEP}
\bibliography{SG}

\end{document}